%% file: main.tex
\pdfoutput=1
\documentclass[11pt]{article}

\IfFileExists{tgpagella.sty}{
  \usepackage{tgpagella}
}{
  \IfFileExists{mathpazo.sty}{\usepackage{mathpazo}}{}
}

\makeatletter
\IfFileExists{lato.sty}{
  \usepackage[defaultsans]{lato}
}{
  \usepackage[scaled=0.95]{helvet}
  
}
\makeatother

\usepackage{amsmath, amssymb}
\usepackage{graphicx}
\usepackage{geometry}
\usepackage{titlesec}
\usepackage{caption}
\usepackage{siunitx}
\usepackage{fancyhdr}
\usepackage{booktabs}
\usepackage{array}
\usepackage{longtable}
\usepackage{multirow}
\usepackage[hidelinks,hypertexnames=false]{hyperref}
\hypersetup{
  pdftitle={Supplementary Information: LLM-Assisted Generation of Transparent, Open-Source Multiphysics Models of Electrochemical Devices},
  pdfauthor={Sebastian Castro, Maya F. Schuchert, Spencer McCluskey, Eric W. Lees, Justin C. Bui}
}
\titleformat{\section}{\sffamily\bfseries\large}{Supplementary Note \thesection}{0.6em}{}
\titleformat{\subsection}{\sffamily\bfseries}{\thesubsection}{0.6em}{}
\titleformat{\subsubsection}{\sffamily\bfseries}{\thesubsubsection}{0.6em}{}

\usepackage{titling}
\pretitle{\begin{flushleft}\sffamily\bfseries\LARGE}
\posttitle{\end{flushleft}\vskip 0.4em}
\preauthor{\begin{flushleft}\rmfamily}
\postauthor{\end{flushleft}\vskip 0.2em}
\predate{\begin{flushleft}\rmfamily}
\postdate{\end{flushleft}\vskip 0.8em}

\renewcommand{\thesection}{\arabic{section}}

\title{LLM-Assisted Generation of Transparent, Open-Source Multiphysics Models of Electrochemical Devices}
\author{Sebastian Castro\\ Department of Chemical and Biomolecular Engineering, New York University, Brooklyn, NY, USA
\and Maya F. Schuchert\\ Department of Chemical and Biomolecular Engineering, New York University, Brooklyn, NY, USA
\and Spencer A. McCluskey\\ Department of Chemical and Biological Engineering, University of British Columbia, Vancouver, BC, Canada
\and Eric W. Lees\\ Department of Chemical and Biological Engineering, University of British Columbia, Vancouver, BC, Canada\\ Corresponding author. e-mail: eric.lees@ubc.ca
\and Justin C. Bui\\ Department of Chemical and Biomolecular Engineering, New York University, Brooklyn, NY, USA\\ Corresponding author. e-mail: justinbui@nyu.edu}
\date{}

\usepackage[T1,OT1]{fontenc}   
\usepackage{textcomp}
\usepackage{latexml}       
\usepackage{xcolor}
\usepackage{float}
\usepackage{fancyvrb}
\usepackage{fvextra}
\usepackage{pdfrender}
\usepackage{pmboxdraw}
\usepackage{accsupp}
\usepackage{calc}
\usepackage{etoolbox}
\hypersetup{
  pdftitle={LLM-Assisted Generation of Transparent, Open-Source Multiphysics Models of Electrochemical Devices},
  pdfauthor={Sebastian Castro, Maya F. Schuchert, Spencer A. McCluskey, Eric W. Lees, Justin C. Bui}
}
\definecolor{nyuviolet}{HTML}{57068C}

\makeatletter
\newcommand{\uchar}[2]{\leavevmode\BeginAccSupp{method=hex,unicode,ActualText=#1}#2\EndAccSupp{}}
\newcommand{\udc}[2]{\DeclareUnicodeCharacter{#1}{\texorpdfstring{\uchar{#1}{#2}}{\unichar{"#1}}}}
\makeatother
\udc{0394}{\ensuremath{\Delta}}   \udc{03A3}{\ensuremath{\Sigma}}   \udc{03A6}{\ensuremath{\Phi}}
\udc{03B1}{\ensuremath{\alpha}}   \udc{03B4}{\ensuremath{\delta}}   \udc{03B5}{\ensuremath{\varepsilon}}
\udc{03B7}{\ensuremath{\eta}}     \udc{03BA}{\ensuremath{\kappa}}   \udc{03BB}{\ensuremath{\lambda}}
\udc{03BC}{\ensuremath{\mu}}      \udc{03BD}{\ensuremath{\nu}}      \udc{03C0}{\ensuremath{\pi}}
\udc{03C1}{\ensuremath{\rho}}     \udc{03C3}{\ensuremath{\sigma}}   \udc{03C4}{\ensuremath{\tau}}
\udc{03C6}{\ensuremath{\varphi}}  \udc{03C9}{\ensuremath{\omega}}
\udc{1D40}{\textsuperscript{T}}   \udc{1D62}{\textsubscript{i}}     \udc{1D63}{\textsubscript{r}}
\udc{2070}{\textsuperscript{0}}   \udc{2074}{\textsuperscript{4}}   \udc{2075}{\textsuperscript{5}}
\udc{2076}{\textsuperscript{6}}   \udc{2077}{\textsuperscript{7}}   \udc{2078}{\textsuperscript{8}}
\udc{2079}{\textsuperscript{9}}   \udc{207A}{\textsuperscript{+}}   \udc{207B}{\textsuperscript{\ensuremath{-}}}
\udc{2080}{\textsubscript{0}}     \udc{2081}{\textsubscript{1}}     \udc{2082}{\textsubscript{2}}
\udc{2083}{\textsubscript{3}}     \udc{2084}{\textsubscript{4}}     \udc{2085}{\textsubscript{5}}
\udc{2090}{\textsubscript{a}}     \udc{2C7C}{\textsubscript{j}}
\udc{2016}{\ensuremath{\|}}       \udc{2190}{\ensuremath{\leftarrow}}  \udc{2192}{\ensuremath{\rightarrow}}
\udc{2194}{\ensuremath{\leftrightarrow}}  \udc{21CC}{\ensuremath{\rightleftharpoons}}
\udc{21D2}{\ensuremath{\Rightarrow}}      \udc{21D4}{\ensuremath{\Leftrightarrow}}
\udc{2202}{\ensuremath{\partial}} \udc{2207}{\ensuremath{\nabla}}   \udc{2208}{\ensuremath{\in}}
\udc{2212}{\ensuremath{-}}        \udc{221A}{\ensuremath{\surd}}    \udc{221D}{\ensuremath{\propto}}
\udc{221E}{\ensuremath{\infty}}   \udc{222B}{\ensuremath{\int}}     \udc{2248}{\ensuremath{\approx}}
\udc{2260}{\ensuremath{\neq}}     \udc{2261}{\ensuremath{\equiv}}   \udc{2264}{\ensuremath{\leq}}
\udc{2265}{\ensuremath{\geq}}     \udc{226A}{\ensuremath{\ll}}      \udc{2272}{\ensuremath{\lesssim}}
\udc{2273}{\ensuremath{\gtrsim}}  \udc{25A1}{\ensuremath{\square}}  \udc{25B3}{\ensuremath{\triangle}}
\udc{25A0}{\ensuremath{\blacksquare}}  \udc{25CF}{\raisebox{0.1ex}{\large\textbullet}}
\udc{2713}{\ensuremath{\checkmark}}    \udc{26A0}{\raisebox{-0.1ex}{\ensuremath{\triangle}}\kern-0.62em\raisebox{0.15ex}{\tiny\bfseries !}\kern0.3em}
\newcommand{\maybeuchar}[2]{#2}
\DeclareUnicodeCharacter{2013}{\texorpdfstring{\maybeuchar{2013}{\textendash}}{\textendash}}
\DeclareUnicodeCharacter{2014}{\texorpdfstring{\maybeuchar{2014}{\textemdash}}{\textemdash}}
\newcommand{\boxuchar}[2]{\leavevmode\BeginAccSupp{method=hex,unicode,ActualText=#1}\rlap{\textpdfrender{TextRenderingMode=Invisible}{-}}#2\EndAccSupp{}}
\newcommand{\bdc}[2]{\DeclareUnicodeCharacter{#1}{\texorpdfstring{\boxuchar{#1}{#2}}{\unichar{"#1}}}}
\bdc{2500}{\textSFx}  \bdc{2502}{\textSFxi}  \bdc{2514}{\textSFii}  \bdc{251C}{\textSFviii}
\DeclareUnicodeCharacter{E000}{\texorpdfstring{\uchar{00630304}{\={c}}}{c}}
\DeclareUnicodeCharacter{E001}{\texorpdfstring{\uchar{00750304}{\={u}}}{u}}

\makeatletter
\newcommand{\currtocext}{toc}
\newcommand{\installparttoc}{%
  \let\orig@addcontentsline\addcontentsline
  \renewcommand{\addcontentsline}[3]{%
    \def\@tmpext{##1}\def\@tmptoc{toc}%
    \orig@addcontentsline{##1}{##2}{##3}%
    \ifx\@tmpext\@tmptoc\ifx\currtocext\@tmptoc\else
      \orig@addcontentsline{\currtocext}{##2}{##3}%
    \fi\fi}}
\AddToHook{begindocument/end}{\installparttoc}
\newcommand{\parttableofcontents}{%
  \section*{\contentsname\@mkboth{\MakeUppercase\contentsname}{\MakeUppercase\contentsname}}%
  \@starttoc{\currtocext}}
\makeatother

\fancypagestyle{msstyle}{%
  \fancyhf{}%
  \fancyhead[L]{\fontencoding{T1}\sffamily\fontsize{9}{11}\selectfont arXiv -- Research Article}%
  \fancyhead[R]{\raisebox{-0.25\height}{\includegraphics[width=0.8247in]{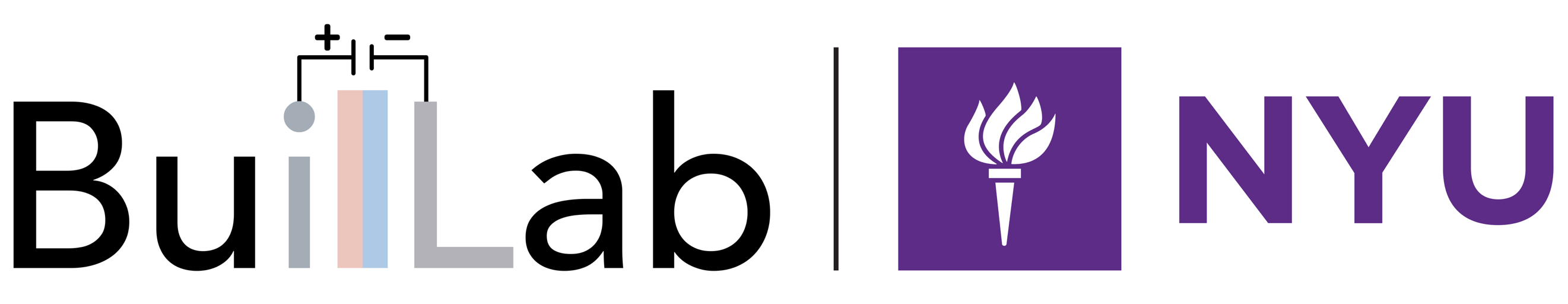}}}%
  \fancyfoot[C]{\fontencoding{T1}\rmfamily\fontsize{9}{11}\selectfont\thepage}}

\newcommand{\msc}[1]{\hyperlink{cite.ms#1}{#1}}

\makeatletter
\newenvironment{msbibliography}{%
  \phantomsection\addcontentsline{toc}{section}{References}%
  \section*{References}%
  \list{\@arabic\c@enumiv.\hfill}{%
    \setlength{\labelwidth}{0.35in}\setlength{\labelsep}{0pt}%
    \setlength{\leftmargin}{0.35in}\setlength{\itemsep}{0pt}%
    \setlength{\parsep}{0pt}\setlength{\topsep}{0pt}%
    \usecounter{enumiv}\let\p@enumiv\@empty
    \renewcommand\theenumiv{\@arabic\c@enumiv}}%
  \sfcode`\.\@m}
  {\def\@noitemerr{\@latex@warning{Empty `msbibliography' environment}}\endlist}
\makeatother
\iflatexml 
  \renewenvironment{msbibliography}{\begin{thebibliography}{44}}{\end{thebibliography}}
\fi

\makeatletter
\def\DejaVuSansMono@scale{0.96}
\newcommand{\guidesizes}{%
  \renewcommand\normalsize{\@setfontsize\normalsize\@xpt\@xiipt
    \abovedisplayskip 10\p@ \@plus2\p@ \@minus5\p@
    \abovedisplayshortskip \z@ \@plus3\p@
    \belowdisplayshortskip 6\p@ \@plus3\p@ \@minus3\p@
    \belowdisplayskip \abovedisplayskip
    \let\@listi\@listI}%
  \renewcommand\small{\@setfontsize\small\@ixpt{11}}%
  \renewcommand\footnotesize{\@setfontsize\footnotesize\@viiipt{9.5}}%
  \renewcommand\scriptsize{\@setfontsize\scriptsize\@viipt\@viiipt}%
  \renewcommand\tiny{\@setfontsize\tiny\@vpt\@vipt}%
  \renewcommand\large{\@setfontsize\large\@xiipt{14}}%
  \renewcommand\Large{\@setfontsize\Large\@xivpt{18}}%
  \renewcommand\LARGE{\@setfontsize\LARGE\@xviipt{22}}%
  \renewcommand\huge{\@setfontsize\huge\@xxpt{25}}%
  \renewcommand\Huge{\@setfontsize\Huge\@xxvpt{30}}%
  \normalsize}
\newcommand{\guideparskip}{%
  \setlength{\parindent}{\z@}%
  \setlength{\parskip}{0.5\baselineskip \@plus 2\p@}%
  \def\@listI{\leftmargin\leftmargini \topsep\z@ \parsep\parskip \itemsep\z@}%
  \let\@listi\@listI \@listi
  \def\@listii{\leftmargin\leftmarginii \labelwidth\leftmarginii \advance\labelwidth-\labelsep \topsep\z@ \parsep\parskip \itemsep\z@}%
  \def\@listiii{\leftmargin\leftmarginiii \labelwidth\leftmarginiii \advance\labelwidth-\labelsep \topsep\z@ \parsep\parskip \itemsep\z@}%
  \pretocmd{\@starttoc}{\parskip\z@}{}{}}
\makeatother

\makeatletter
\patchcmd\longtable{\par}{\if@noskipsec\mbox{}\fi\par}{}{}
\makeatother

\DefineVerbatimEnvironment{Highlighting}{Verbatim}{commandchars=\\\{\}}

\begin{document}

\fontencoding{T1}\renewcommand{\encodingdefault}{T1}%
\linespread{1.096}\normalfont\normalsize
\setcounter{secnumdepth}{0}
\titleformat{\section}{\sffamily\bfseries\fontsize{12}{14.5}\selectfont\color{nyuviolet}}{}{0pt}{}
\titleformat{\subsection}{\sffamily\bfseries\fontsize{12}{14.5}\selectfont\color{nyuviolet}}{}{0pt}{}
\titlespacing*{\section}{0pt}{18pt}{8pt}
\titlespacing*{\subsection}{0pt}{12pt}{6pt}
\setlength{\parindent}{0pt}\setlength{\parskip}{6pt plus 1pt}
\hyphenpenalty=10000 \exhyphenpenalty=10000 \tolerance=3000 \emergencystretch=3em
\captionsetup{labelfont=bf,labelsep=period,justification=justified,singlelinecheck=false,font=normalsize,skip=6pt}
\captionsetup[figure]{name=Fig.}
\urlstyle{same}
\pagestyle{msstyle}
\input{ms}

\end{document}

%% file: ms.tex
\iflatexml\else 
\begingroup\setlength{\parskip}{0pt}%
{\sffamily\bfseries\fontsize{12}{14.5}\selectfont\color{nyuviolet}%
LLM-Assisted Generation of Transparent, Open-Source\par\vspace{2pt}
Multiphysics Models of Electrochemical Devices\par\vspace{2pt}}
\mbox{}\par
Sebastian Castro\textsuperscript{1}, Maya F. Schuchert\textsuperscript{1}, Spencer A. McCluskey\textsuperscript{2}, Eric W. Lees\textsuperscript{2,*}, Justin C. Bui\textsuperscript{1,*}\par
\mbox{}\par
\textsuperscript{1}Department of Chemical and Biomolecular Engineering, New York University, Brooklyn, NY, USA\par
\textsuperscript{2}Department of Chemical and Biological Engineering, University of British Columbia, Vancouver, BC, Canada\par
\textsuperscript{*}Corresponding authors. e-mail: justinbui@nyu.edu; eric.lees@ubc.ca\par
\endgroup
\clearpage
\fi

\section{Abstract}\label{ms:abstract}

Multiphysics continuum models are powerful tools for studying electrochemical devices, enabling in silico reactor design and resolution of local pH, potential, and concentration fields that govern device performance but are difficult to measure experimentally. However, constructing such models requires substantial numerical expertise or reliance on proprietary software. Here, we show that frontier large language model agents can remove this implementation burden while keeping the underlying physics under researcher control. Using one-dimensional electrochemical CO₂ reduction to CO in a porous gas diffusion electrode as a test case, we develop a machine-readable, human-specified modeling harness containing governing equations, parameters, numerical methods, logical build stages, and human-verifiable checkpoints. From this specification, the agent reproducibly constructs complete multiphysics models in open-source Julia. Independently built models, including fully autonomous agent-built models, agree with an equivalent COMSOL implementation to within 0.7\% of the peak CO partial current density, and with one another to within 0.04\%. Systematically planted errors demonstrate the importance of explicit specifications for reproducibility and reveal the agent's capabilities and limitations in debugging model physics. This framework establishes a more transparent approach to multiphysics modeling in which physical descriptions and governing equations, rather than specialized code, become the primary inputs for computational model development.

\clearpage
Electrochemical reactors sit at the center of the energy transition: electrolyzers use renewable electricity to drive the synthesis of fuels and chemicals, while fuel cells and batteries store that renewable electricity in chemical form\textsuperscript{\msc{1},\msc{2}}. Each of these devices is a coupled, multi-component system for which performance is an emergent property of the interactions between material components (electrodes, membranes, and electrolytes) and operating conditions (temperature, pressure, flow rate). The spatially heterogeneous pH, potential, and reactant concentration profiles within these reactors directly govern selectivity, throughput, and durability, but these local chemical environments are largely inaccessible to direct measurement under realistic operating conditions\textsuperscript{\msc{3},\msc{4}}. Consequently, rational reactor design and optimization are difficult to achieve through experimentation alone.

Continuum modeling can accelerate the optimization of electrochemical devices by translating their coupled physics (\emph{i.e.,} multiphysics) into a system of governing equations that can be solved numerically\textsuperscript{\msc{5},\msc{6}}. Conservation equations for mass, charge, and momentum, expressed with models of ion transport, electrochemical kinetics, and fluid flow in porous media, capture the relationship between local chemical environments and experimental observables. For instance, multiphysics models can resolve how competing reactions distribute current across an electrode\textsuperscript{\msc{7}}, how membrane chemistry impacts voltage losses\textsuperscript{\msc{8}}, and how spectator ions tune reaction selectivity\textsuperscript{\msc{9}}. Once parameterized and validated against experimental data, they also enable large design spaces to be explored at a fraction of the cost of experiments.

The history of electrochemical continuum modeling is a history of shifting barriers. In the 1960s, building a model required simultaneous mastery of the physics of electrochemistry and the numerical methods needed to solve the resulting stiff, coupled partial differential equations (PDEs), often using Fortran\textsuperscript{\msc{10},\msc{11}}, which confined serious modeling to a small community. Commercial simulation platforms, principally COMSOL Multiphysics, placed those algorithms behind a graphical interface and a proprietary solver\textsuperscript{\msc{12}}, allowing practitioners to focus more effort on defining the physical equations and less on implementation. Physical intuition remained essential to choosing and evaluating the governing equations, constitutive relationships, and boundary conditions; the numerical machinery, however, receded into the background, and with it direct engagement with how models are solved. When a solver fails, diagnosing whether the failure originates from the governing equations, discretization, or numerical implementation can therefore be difficult. Generic messages such as ``maximum iterations reached'' rarely indicate which equation, term, or mesh element is responsible, leaving debugging to heuristic experience with mesh refinement, relaxation factors, and initial guesses. Commercial solvers are also less accessible and more expensive than open-source alternatives\textsuperscript{\msc{13},\msc{14}}, and proprietary formats complicate reproducibility, as model files are often tied to specific software versions and inaccessible to collaborators without licenses\textsuperscript{\msc{15}}.

We contend that the emergence of large language model (LLM) agents is the next chapter in electrochemical continuum modeling. A researcher with deep domain knowledge but limited numerical training can instruct an artificial intelligence (AI) agent to implement a fully open-source continuum model in which every equation, solver step, and line of code is visible, modifiable, and explainable. The agent does not replace the physics expertise of the researcher but rather replaces the implementation burden. The researcher must still specify the governing equations precisely, anticipate which physical mechanisms are relevant, design staged verification checks that catch errors before they propagate, and recognize when a converged solution is unphysical. Domain knowledge remains critical; what changes is that the auxiliary technical burden shifts from Fortran subroutines and COMSOL debugging toward specification and verification. Because the resulting model is open-source and every step of its numerics is inspectable, the practical barrier to continuum multiphysics modeling becomes the ability to state the physics precisely rather than access to a solver license or extensive training in numerical methods.

Agentic AI has already shown potential for integration into the scientific process, including autonomy in designing experiments, executing code, and discovering new pathways for chemical synthesis\textsuperscript{\msc{16},\msc{17}}. LLM agents have been shown to autonomously write, execute, and debug finite-element code for continuum mechanics problems\textsuperscript{\msc{18}}, collaborate as multi-agent teams across a range of physics domains\textsuperscript{\msc{19},\msc{20}}, and generate PDE solvers that match expert-written code on benchmarks including compressible Navier--Stokes\textsuperscript{\msc{21}}. Nonetheless, this work has largely targeted weakly coupled or single-physics problems, leaving strongly coupled multiphysics systems comparatively unexplored. A recent multi-agent framework, for instance, runs finite-element simulations from natural language by retrieving domain knowledge from a curated knowledge graph and verifying the solver against closed-form analytical solutions\textsuperscript{\msc{22}}. Neither approach transfers directly here: a nonlinear, strongly coupled system of this kind generally possesses no analytical reference solution, while its governing physics must be specified through researcher-authored modeling choices specific to the simulated device rather than retrieved as a fixed set of facts. Electrochemical reactors are a particularly demanding case, with acid--base equilibria, interfacial kinetics, migration, and multicomponent gas transport entering one another\textquotesingle s source terms and transport coefficients. The equations must be solved simultaneously as a single nonlinear system, with buffer equilibria that relax orders of magnitude faster than species diffuse, making that system stiff and difficult to solve robustly.

\begin{figure}[t!]
\centering
\includegraphics[width=\textwidth]{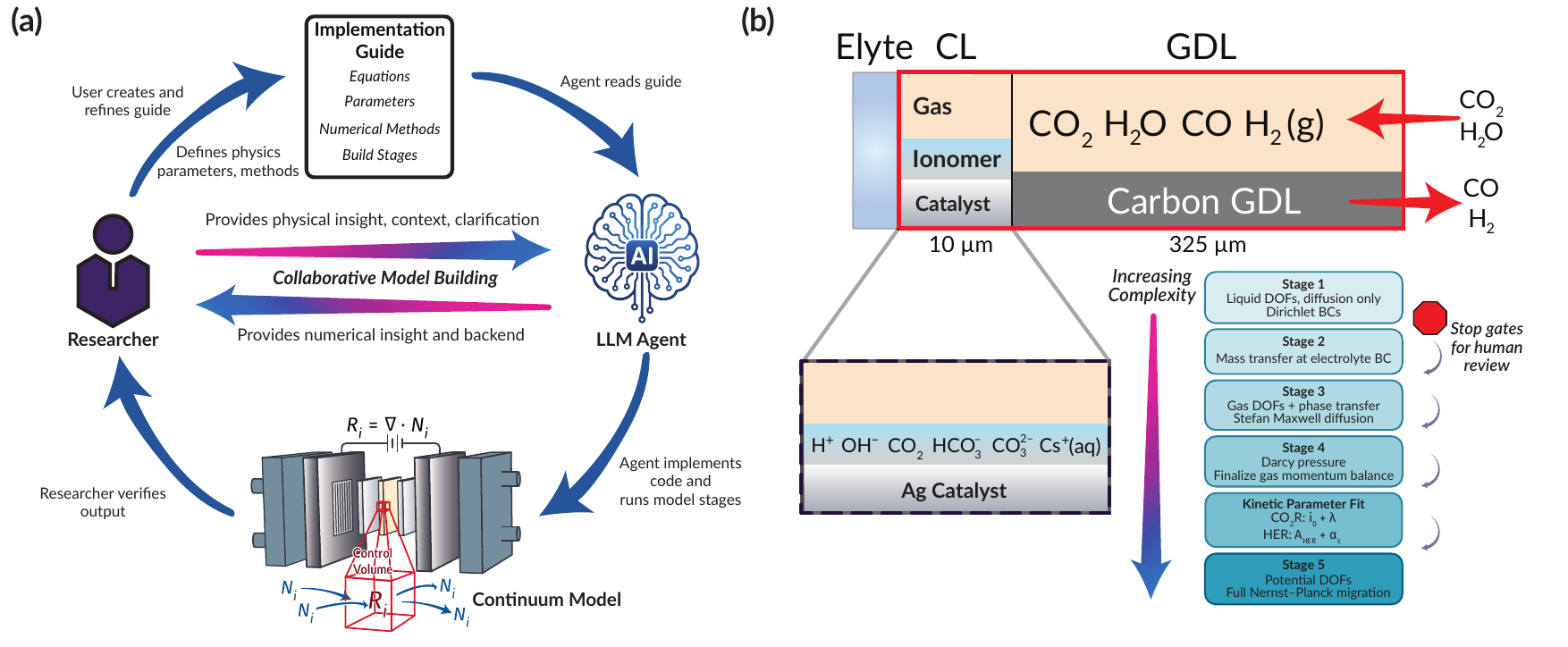}
\caption{\textbf{(a)} Schematic of the agentic multiphysics modeling workflow employed here with a human-in-the-loop. \textbf{(b)} Schematic of the CO₂ reduction porous electrode model case study, along with the 5-stage implementation approach in Julia, with the primary physics additions/changes included in each stage. Elyte, electrolyte; CL, catalyst layer; GDL, gas diffusion layer; DOFs, degrees of freedom; BC, boundary condition; HER, hydrogen evolution reaction.}\label{fig:1}
\end{figure}

To demonstrate the validity of our approach, we developed a detailed, machine-readable implementation guide for a one-dimensional (1D) continuum model (\hyperref[fig:1]{\textbf{Fig. 1}}) of silver-catalyzed CO\textsubscript{2} reduction (CO\textsubscript{2}R) in a gas diffusion electrode (GDE). This implementation guide served as a harness that directed the AI agent through construction of the multiphysics model in open-source Julia\textsuperscript{\msc{23}}. Researchers of varying experience levels used the guide to instruct AI agents to build the model, without writing a line of code themselves. Because the complex numerical system has no analytical solution, validation of the approach was demonstrated against an equivalent model developed in COMSOL. Reproducibility was assessed by comparing independently generated model implementations across users. Beyond accuracy and reproducibility, the agentic workflow provides a transparent interface to the numerical implementation: the agent that built the code can inspect the Jacobian, trace convergence failures, and explain the numerical behavior in the language of physics directly to a human operator. Collectively, this framework establishes an approach to multiphysics modeling in which human-specified equation sets and natural-language instructions, rather than specialized code and numerical implementations, become the primary inputs for model development. In this paradigm, the specification harness---rather than any individual implementation---becomes the central scientific artifact that other practitioners can audit, reproduce, modify, and build upon.

\section{Results}\label{ms:results}

\subsection{A specification-driven, staged build}\label{ms:a-specification-driven-staged-build}

The modeling of electrochemical reactors is a prime example of a complex, multiphysics problem, and thus presents a suitable platform for establishing agentic multiphysics approaches. Every one of the nonlinear couplings introduced above, and the corresponding numerical strategy that makes the resulting stiff system tractable, must be stated explicitly before an agent can implement it. Hence, an implementation guide (\emph{i.e.,} machine-readable specifications, here markdown files) can serve as a harness for an agent building multiphysics models, fully describing the problem to be modeled: all relevant parameters, equations, and boundary conditions, together with a scaffolded build plan that the coding agent must follow. These guides link the scientific knowledge and intuition from the human to the numerical implementation of the agent. Precise and comprehensive specifications ensure consistency and accuracy between agents with probabilistic behavior, and the responsibility remains with the researcher to accurately define the problem. In other words, the agent\textquotesingle s primary role in this workflow is the implementation of a specified guide, rather than the inference of a complete model from an incomplete specification.

To provide the agent (and its human operator) with logical checkpoints at which to assess and correct mistakes in implementation, the guide decomposes model development into five stages of increasing physical complexity (\textbf{\hyperref[fig:1]{Fig. 1}b} and Supplementary Fig. 1), from liquid Nernst--Planck transport, buffer reactions, and electrochemical kinetics at 400 degrees of freedom (DOFs) to the full electrostatically coupled system at 1,480 DOFs on the production mesh. Between stages, stop gates allow the human-in-the-loop to review quantitative physicality checks and diagnostic plots prescribed by the guide itself, rather than leaving them to the agent\textquotesingle s discretion, before authorizing the agent to advance to the next stage. These gates are solution-verification checks in the established sense\textsuperscript{\msc{24}}, applied by the human rather than the agent. Full stage definitions, pass/fail criteria, and details of the fitting step used to parameterize the model against the experimental data of Romiluyi et al.\textsuperscript{\msc{25}} are given in the Methods section.

\begin{figure}[b!]
\centering
\includegraphics[width=\textwidth]{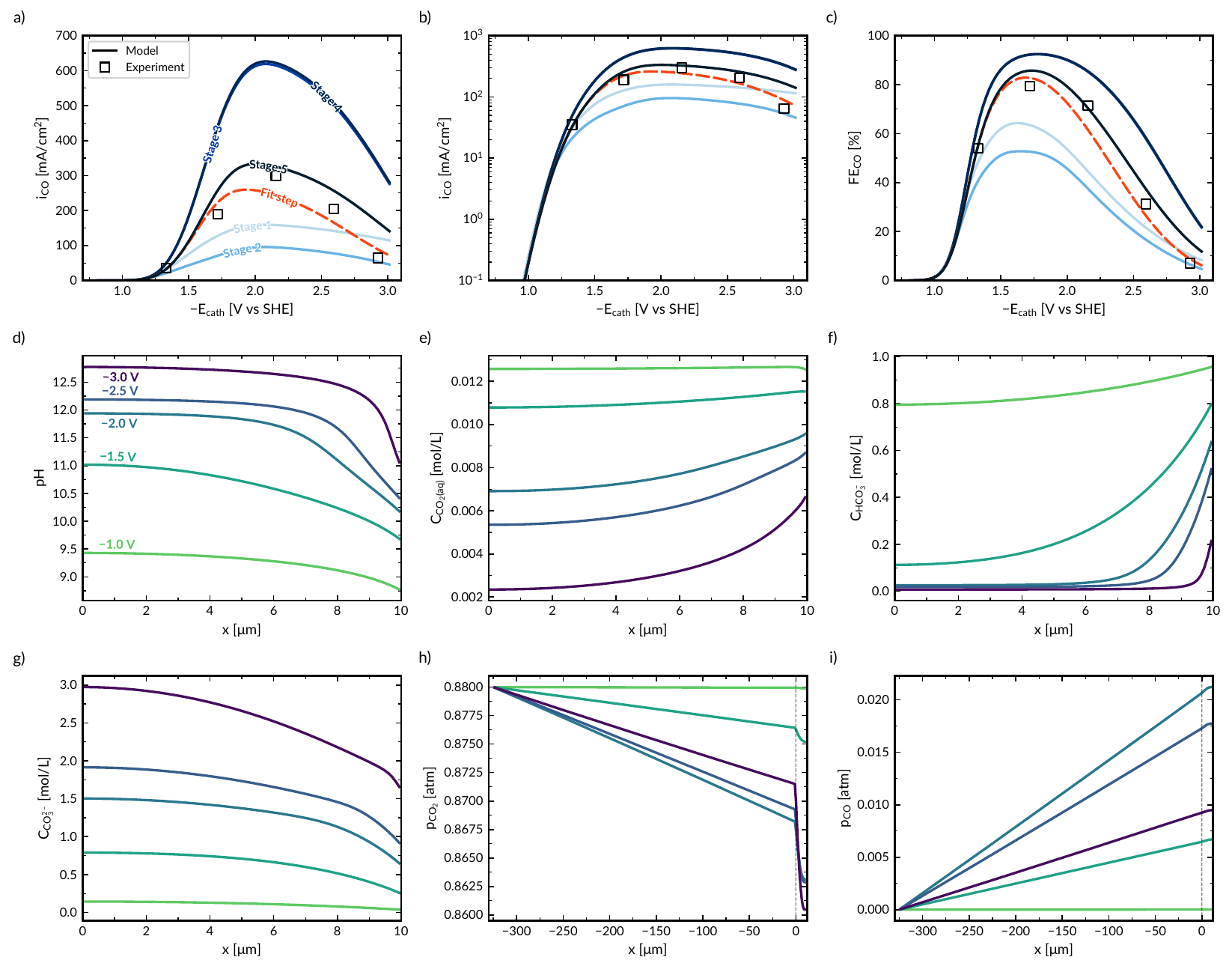}
\caption{\textbf{Stage-by-stage build progression and internal fields of final model.} CO \textbf{(a)} linear-scale polarization, \textbf{(b)} log-scale polarization and \textbf{(c)} faradaic efficiency after each build stage and the fit. Converged Stage-5 profiles at five applied potentials: across the catalyst layer, \textbf{(d)} pH, \textbf{(e)} CO\textsubscript{2}, \textbf{(f)} HCO\textsubscript{3}\textsuperscript{--} and \textbf{(g)} CO\textsubscript{3}\textsuperscript{2--} concentrations; across the gas diffusion layer and catalyst layer, \textbf{(h)} CO\textsubscript{2} and \textbf{(i)} CO partial pressures. Markers represent experimental data. Voltages vs the standard hydrogen electrode (SHE).}\label{fig:2}
\end{figure}

Our assessment of the AI-assisted build begins with the polarization and faradaic efficiency (FE) results, and how these outputs evolve as each stage of the implementation guide adds a layer of physics (\textbf{\hyperref[fig:2]{Fig. 2}a--c)}. Notably, the plots reproduce the canonical behavior of a CO\textsubscript{2} electrolyzer: \emph{FE}\textsubscript{CO} peaks at intermediate potentials and declines rapidly at high potentials where the hydrogen evolution reaction (HER) outcompetes. Physically, this occurs because of the interplay between CO\textsubscript{2} reduction kinetics, CO\textsubscript{2} mass transport, and the consumption of CO\textsubscript{2} by acid--base equilibria within the porous electrode\textsuperscript{\msc{26}}. The evolution of the curves coincides with the physics added at each stage: Stage 1 includes buffer equilibria, electrochemical kinetics, and liquid-phase species transport by diffusion alone, with fixed-concentration (Dirichlet) boundary conditions; the Stage 2 mass-transfer boundary condition allows CO\textsubscript{2} depletion and pH buildup and correspondingly reduces CO partial current, \emph{i}\textsubscript{CO}; Stage 3 phase transfer from the gas adds a liquid-phase CO\textsubscript{2} source and drives \emph{i}\textsubscript{CO} back up; Stage 4 momentum balance in the gas phase adds a small but physical correction; the fit, performed on the Stage 4 model, brings \emph{i}\textsubscript{CO} down to the experimental data; and Stage 5 migration raises \emph{i}\textsubscript{CO} to its final value by increasing the flux of OH\textsuperscript{--} and other anions away from the cathode, which mitigates the salting-out of CO\textsubscript{2}.

Beyond examining the polarization data, the correctness of the implemented physics was complementarily verified at each stage by analyzing the species profiles to confirm the simulated device concentration fields are consistent with physical intuition. The most relevant field profiles are shown in \textbf{\hyperref[fig:2]{Fig. 2}d--i,} and the full output of all the plotted profiles from every stage is shown in Supplementary Figs. 3--13. As expected, the simulated pH (\textbf{\hyperref[fig:2]{Fig. 2}d}) is highest at the gas diffusion layer (GDL) side (x = 0) where the CO\textsubscript{2} consumption rate is highest, generating OH\textsuperscript{--} anions via CO\textsubscript{2}R. Additionally, the dissolved CO\textsubscript{2} (\textbf{\hyperref[fig:2]{Fig. 2}e}) is lowest near the GDL where CO\textsubscript{2}R rates are highest and increases toward the electrolyte, where CO\textsubscript{2} is slightly replenished by incoming (bi)carbonates and their equilibria. The (bi)carbonate field profiles (\textbf{\hyperref[fig:2]{Fig. 2}f--g}) behave in accordance with their acid--base equilibria and CO\textsubscript{2} concentrations. In the gas phase (\textbf{\hyperref[fig:2]{Fig. 2}h--i}), gaseous CO\textsubscript{2} is highest where it comes in from the flow channel, has a linear concentration gradient through the GDL where it undergoes diffusive transport only, and is consumed in the catalyst layer (CL). CO is generated in the CL and diffuses out into the gas channel. These profiles, which would be difficult to measure experimentally in the microscale porous electrode, show that the model is physically reasonable.

\subsection{Agreement with COMSOL and across builds}\label{ms:agreement-with-comsol-and-across-builds}

To assess the efficacy of the specification guide in facilitating agentic multiphysics modeling, we leveraged the build to develop an experimentally calibrated continuum multiphysics model of a gas diffusion electrode used for CO\textsubscript{2}R. We consider an AI-assisted build successful if it yields consistent and physically reasonable results across independent builds, reproduces the simulation results of an established commercial tool reference (in this case, COMSOL) to within a few percent, and remains fully transparent end-to-end (\emph{i.e.,} all code is generated and available for examination by human evaluation or the audits of other agents).

\begin{figure}[t!]
\centering
\includegraphics[width=\textwidth]{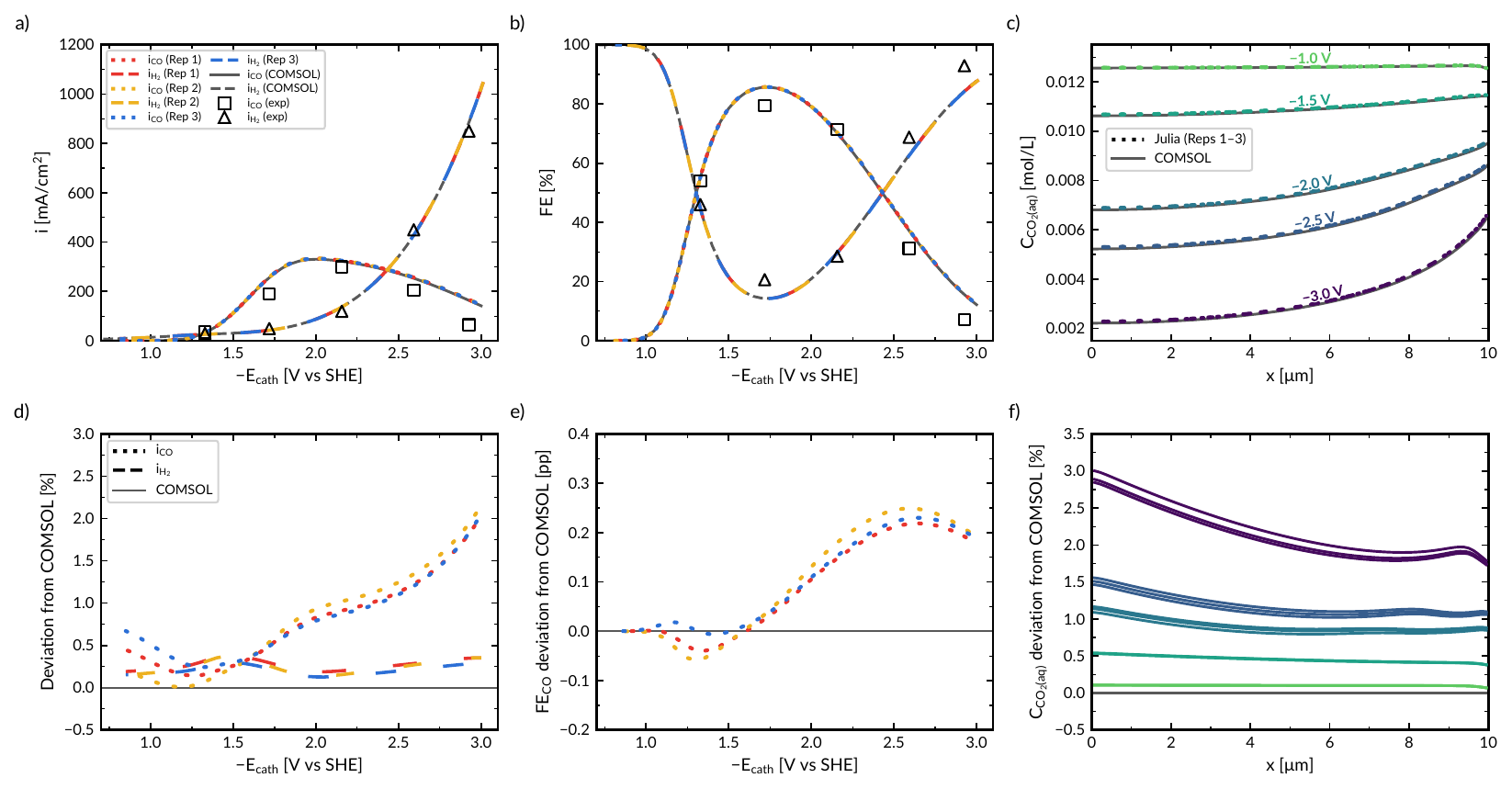}
\caption{\textbf{Agreement of three independently built Julia models with the COMSOL model.} \textbf{(a)} CO and H\textsubscript{2} partial current densities, \textbf{(b)} faradaic efficiencies and \textbf{(c)} CO\textsubscript{2}(aq) concentration across the catalyst layer at five applied potentials, with the deviation of each from COMSOL below it: \textbf{(d)} currents, \textbf{(e)} FE\textsubscript{CO}, in percentage points, and \textbf{(f)} CO\textsubscript{2}(aq). Current and FE deviations are evaluated at the COMSOL sweep potentials. Markers represent experimental data.}\label{fig:3}
\end{figure}

To confirm the accuracy of the agent's multiphysics implementation, we benchmarked the three independent human-in-the-loop builds, made with Claude Opus 5, against a physically equivalent COMSOL model with the same parameters \textbf{(\hyperref[fig:3]{Fig. 3})}. The Julia build outputs agree with the COMSOL model quantitatively, with the polarization and FE curves overlapping entirely. Evaluated at the 44 COMSOL sweep potentials that fall within the Julia sweep window, the average of the root mean square deviations (RMSDs) of each build against COMSOL is 2.27 mA cm\textsuperscript{−2} for \emph{i}\textsubscript{CO}, 1.07 mA cm\textsuperscript{−2} for the H\textsubscript{2} partial current density, \emph{i\textsubscript{H2}}, and 0.14 percentage points for \emph{FE\textsubscript{CO}}. The \emph{i}\textsubscript{CO} RMSD represents 0.7\% of the peak CO current density of about 330 mA cm\textsuperscript{−2}, and the Julia \emph{i}\textsubscript{CO} values sit systematically above the COMSOL values, by up to 2.2\% at currents above 5 mA cm\textsuperscript{−2} and about 0.9\% at the peak, a discrepancy likely due to slight differences in mesh generation or to numerical implementation differences between the solvers.

Across the three builds evaluated on a common 89-point voltage sweep, the partial current densities also agreed with one another to within a pooled RMSD of 0.12 mA cm\textsuperscript{−2} for CO and 0.11 mA cm\textsuperscript{−2} for H₂, and the CO faradaic efficiency agreed within 0.01 percentage points. The fitting approach was also remarkably consistent across the three builds. The CO\textsubscript{2}R Marcus--Hush--Chidsey\textsuperscript{\msc{27}} kinetic parameters were fit to literature data using a Nelder--Mead algorithm\textsuperscript{\msc{28},\msc{29}}, landing on an exchange current density (\emph{i}\textsubscript{0}) of 106.47 ± 0.11 A m\textsuperscript{−2} and a reorganization energy (\emph{λ}) of 1.6111 ± 0.0004 eV. The small residual spread reflects the flatness of the fit objective; all three replicates used identical termination criteria and reached the same sum of squared errors to four significant figures (0.05608). These results confirm that independent implementations of the specification harness produce nearly identical multiphysics models both to each other and to an established commercial software package, providing confidence in both the methodological approach and the accuracy of the physical implementation.

\subsection{Solver implementation and agent cost--benefit deviation}\label{ms:solver-implementation-and-agent-costbenefit-deviation}

\begin{figure}[b!]
\centering
\includegraphics[width=\textwidth]{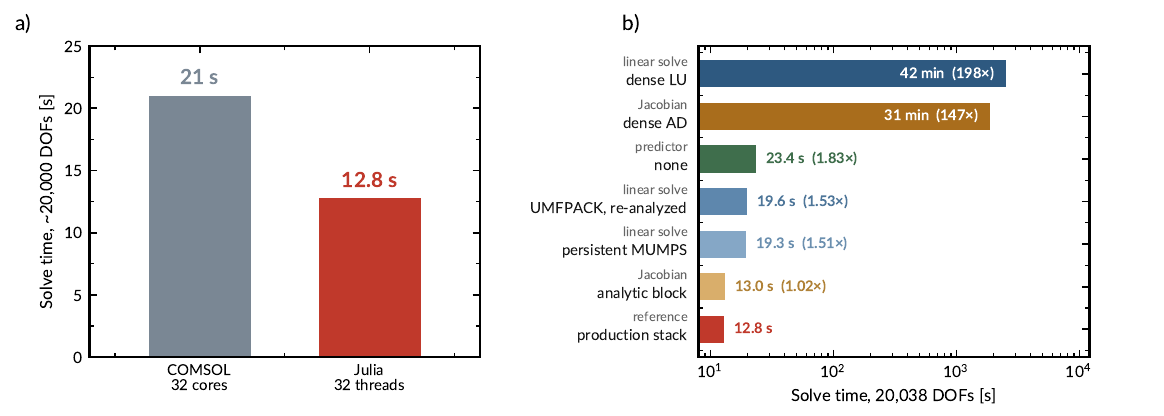}
\caption{\textbf{Solver performance of the Julia implementation. (a)} Solve time for each implementation\textquotesingle s full sweep (89 potentials in Julia, 61 in COMSOL), both run on the same machine at matched core budget (32 cores/threads, ≈20,000 DOFs). \textbf{(b)} Measured cost of Jacobian and linear-solver choices at 20,038 DOFs. Every configuration returns the same polarization curve to the precision written, so bars differ only in cost. Ratios are relative to the configuration specified in the guide (UMFPACK with pattern reuse and forward-mode automatic differentiation). LU, lower--upper factorization; AD, automatic differentiation.}\label{fig:4}
\end{figure}

One benefit of commercial packages is that their plug-and-play numerical solvers have been long-optimized to ensure that solvers are properly matched to the systems that they solve. To benchmark the quality and speed of the agent's numerical implementation of CO\textsubscript{2}R against that of a commercial software package, we performed speed benchmarking against an equivalent COMSOL model with matched degrees of freedom. Notably, the agent's Julia model demonstrates competitive solve speed to COMSOL with the added benefit of customization. Timed on the same machine at a matched compute budget and problem size (\textasciitilde20,000 DOFs), the full 89-potential sweep takes about 13 s with our production solver stack compared to 21 s for COMSOL's 61-potential sweep (\textbf{\hyperref[fig:4]{Fig. 4}a}). It is important to emphasize, however, that a direct comparison is not possible because of inherent differences in discretization (finite element method versus finite volume method), linear-solve library (PARDISO\textsuperscript{\msc{30}} vs UMFPACK\textsuperscript{\msc{31}}), and solve environment (native Windows vs the Windows Subsystem for Linux). The solver ablation (\textbf{\hyperref[fig:4]{Fig. 4}b}) reveals that speed gains in numerical solvers often come from simply tailoring the solver (\emph{e.g.,} its sparsity, the discretization method, \emph{etc.}) to the problem solved. Agentic multiphysics implementations with frontier agents therefore benefit significantly from the cheapness of customizing solver parameters to the specific problem being solved. The gains come almost entirely from exploiting model sparsity, first in the linear solve and again in Jacobian assembly, each worth about two orders of magnitude against a dense alternative. We note that the components of the developed solver do not represent novel mathematics or solution techniques. However, with an open-source, modifiable solver, we no longer need to sacrifice performance for the solver generality that is often required of one-size-fits-all software packages. Full details of the solver stack are given in Supplementary Note 4.

Despite agents mostly conforming to the specification, we observed deliberate, documented deviations during intermediate solver development. Each solver component was explicitly mandated in a reproducibility checklist, yet none of the builds implemented the analytical block Jacobian or installed MUMPS\textsuperscript{\msc{32}}, the open-source sparse direct solver that an earlier version of the guide specified, unless explicitly told to do so for speed benchmarking. The guide itself labeled these components performance multipliers, and the agents took the label at face value, opting for simpler alternatives and recording the reasoning in their build dialogue notes. The agent would weigh, for example, hand-deriving the analytical block Jacobian, which can yield silent calculation errors, against a forward-differentiation call\textsuperscript{\msc{33}} that takes a few lines of code and is correct to machine precision. Automatic differentiation has been adopted for the same reason in Newman-tradition porous electrode codes\textsuperscript{\msc{34}}, where hand-derived Jacobians are a known source of silent error. Similarly, the agents opted for the UMFPACK solver available in the Julia standard library, justifying the substitution as having similar performance while avoiding the extra effort required for installing MUMPS. Regardless, the difference in solve time for the solver items substituted by the agent was non-detrimental (\textbf{\hyperref[fig:4]{Fig. 4}b}), while the implementation burden for the agent would have been intensive. Because both substitutions were at least as fast as the specified components, they were adopted in the final version of the solver stack. This behavior is nonetheless a cautionary tale for other users of this methodology: unless explicitly told and checked by a human operator, frontier agents often route numerics through simpler implementations that they deem equivalent and more efficient.

\subsection{Diagnostic transparency during model development}\label{ms:diagnostic-transparency-during-model-development}

\begin{figure}[H]
\centering
\includegraphics[width=5.9833in]{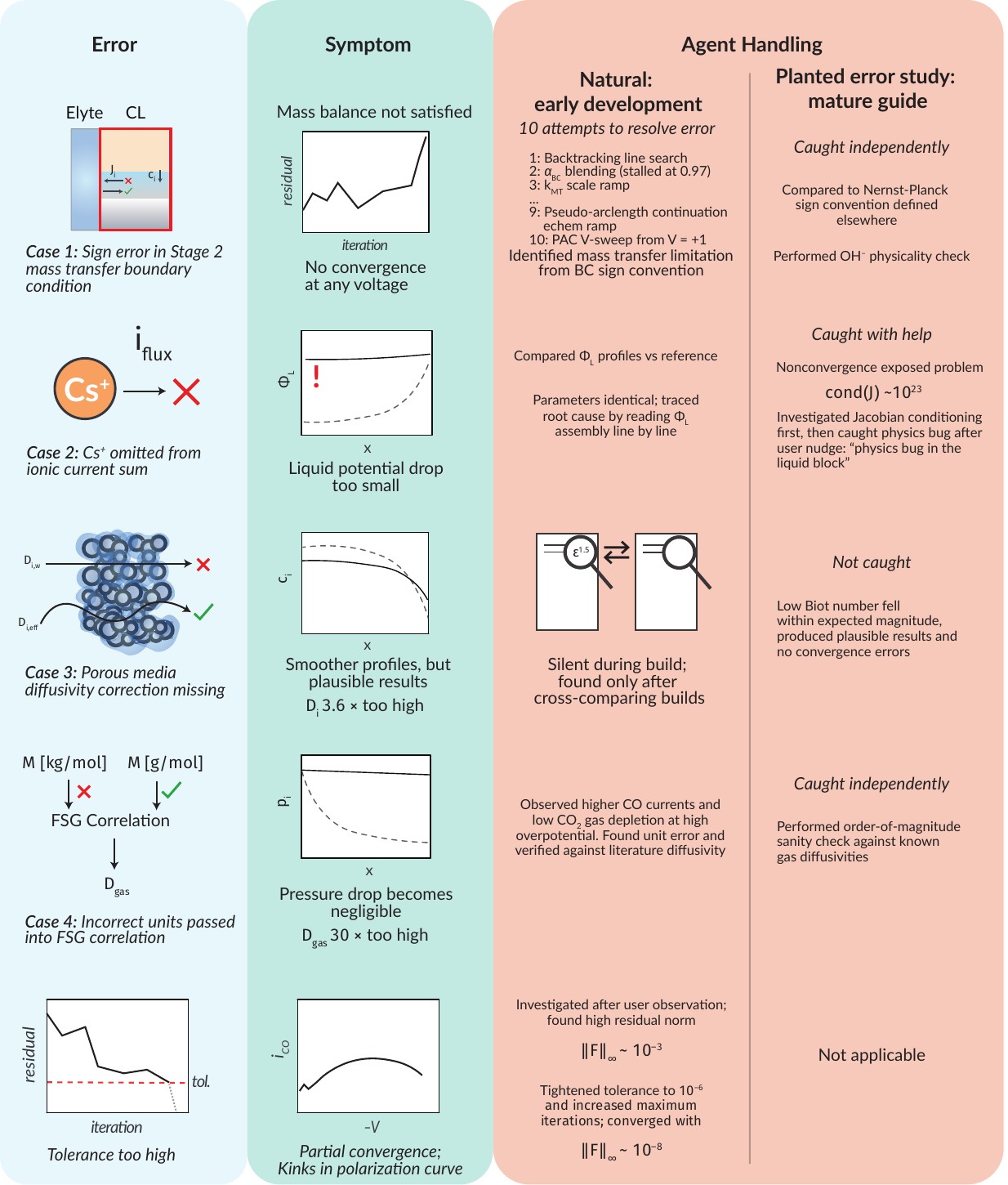}
\caption{\textbf{How does the multiphysics agent handle errors?} Errors encountered naturally during model development and planted systematically into a mature guide, their respective symptoms that the human can see and report to the agent, and the agent's response to resolve the issue. The planted error study was conducted with Claude Opus 4.8. Agents were instructed not to consult outside resources, including other model builds, unless stated otherwise (see Supplementary Notes 6 and 8). Elyte, electrolyte; CL, catalyst layer; BC, boundary condition; α\textsubscript{BC}, boundary-condition blending factor; \emph{k}\textsubscript{MT}, mass-transfer coefficient; PAC, pseudo-arclength continuation; FSG, Fuller--Schettler--Giddings; Φ\textsubscript{L}, liquid-phase potential; cond(J), condition number of the Jacobian.}\label{fig:5}
\end{figure}

Building a correct model depends first on the front-end work of developing and specifying the relevant physics and numerical methods within the implementation guide. Once the guide is sufficiently specified and could generate models, the process consisted of building, troubleshooting, refining, and iterating on the model harness. Intermediate builds using earlier versions of the implementation guide surfaced several critical errors, required substantial human input, and/or produced significant variation in build trajectories even from the same guide (Supplementary Note 7), reflecting the importance of the user in adequately specifying the model physics as to reduce variance in the LLM\textquotesingle s probabilistic output\textsuperscript{\msc{35}}. As the guide was refined, these issues were progressively designed out and constrained, so that the final replicates required only minimal human intervention and had no critical failures. The challenges encountered throughout the development process for the implementation guide, along with the agents' capacity to detect and address these failures, are shown in \hyperref[fig:5]{\textbf{Fig. 5}}.

An example of a physics bug that arose in the early builds occurred when switching from the Dirichlet to the Neumann mass-transfer boundary condition for CO\textsubscript{2} in Stage 2, where the model suddenly could not converge. The agent reasoned that the mass-transfer boundary condition was unable to sustain high currents, and, upon investigating the physics, found a single sign error in the mass-transfer flux term as the root cause, since it incorrectly caused CO\textsubscript{2} to leave the system rather than being supplied from the electrolyte side. The same sign error recurred independently in five of six early builds, across two agent models and four users, because the guide gave the flux without properly specifying its sign convention. In all runs the solver failed to converge, prompting investigation. In some cases, the agent found the bug as discussed above, while in others it could not identify the root cause without human intervention (\emph{e.g.,} prompting the agent to check the sign on the boundary condition).

Even when the physics is implemented correctly, convergence errors are common in numerical models and are the most common culprits that stall model development. In two independent late-stage builds from different users, spurious points in the polarization curve appeared that partially converged but were not on the correct solution path. In both cases, the human-in-the-loop identified the issue through visual plot inspection and communicated it to their agent, which found that loose convergence criteria were admitting partially converged points. The agents autonomously converged on the same fix, tightening the absolute residual tolerance.

Nonetheless, the agents struggled to identify hidden physics bugs that do not block convergence, requiring human input to eliminate these physical inaccuracies. These silent errors are the most dangerous, since they can yield results that appear reasonable while being inaccurate. For example, an early guide included two expressions for the effective diffusivity of liquid-phase species, one with the Bruggeman correction for porous media and one without, resulting in effective diffusivities that differed by a factor of 3.54. In this case, none of the agents mentioned the ambiguity. Fortunately, when the agent was then instructed to cross-compare two replicates that had significant deviations in CO partial current density, it identified the difference in the parameter definition line by line. Notably, the ease of comparison was made possible by the standardization of the modular codebase in the specification guide. The agent could explain the physical implications of the differences in CO\textsubscript{2} diffusion coefficients, tracing the user\textquotesingle s observation of higher CO partial currents to the fitting optimizer compensating for the decreased diffusivity by increasing exchange current density. It also located the two conflicting definitions in the guide itself regarding the use of the Bruggeman correction, which were reconciled in the next iteration of the specification.

Because the agent writes every line of the solver, it does not merely return a number, exit flag, or convergence message; it can narrate the numerics in the language of the physics. During our builds, the agent could inspect the 1,480×1,480 sparse Stage-5 Jacobian and report its condition number, identify that the cell at the gas-side boundary carried the highest CO₂R reaction rate while the electrolyte-side cell was HER-dominated, and explain why in terms of local CO₂ availability and pH, and verify current continuity against the boundary flux. This diagnostic ability represents a key benefit of having a specialized agent for each multiphysics problem as opposed to relying on a generalized software package. When using commercial software, errors can be difficult to self-identify, leading to reliance on reporting to a dedicated support team. Conversely, agents can point to specific variables and their residuals in a given mesh element, trace that to the exact physics, and tune solvers for convergence barriers such as stiffness and near-zero concentration terms.

\subsection{Error handling with multiphysics agents}\label{ms:error-handling-with-multiphysics-agents}

To test the agent\textquotesingle s ability to detect and resolve issues in a controlled manner and establish the robustness of the guide itself, we conducted a planted-error study. Unlike the natural errors above, which were a result of and could be compounded by guide incompleteness or ambiguity, these were systematically placed in separate copies of a mature pre-release version of the guide to assess the model's ability to identify and handle these errors. For four planted errors (referred to as Cases 1--4), the agent was not told which relation had been modified and was prompted to build without referring to any files besides the specification, just as any other build. A fifth build used the unmodified guide as a control. All builds used Claude Opus 4.8 at maximum effort and received the same prompt. The four defects included a sign reversal in a boundary condition (Case 1, \textbf{\hyperref[fig:5]{Fig. 5}} row 1), an omitted term in a conservation law (Case 2, \textbf{\hyperref[fig:5]{Fig. 5}} row 2), an omitted correction factor in a constitutive relation (Case 3, \textbf{\hyperref[fig:5]{Fig. 5}} row 3), and a unit inconsistency in an empirical correlation (Case 4, \textbf{\hyperref[fig:5]{Fig. 5}} row 4). Each reproduces an error encountered naturally during development rather than one invented for the study (Supplementary Note 7). However, sandboxing these errors intentionally provides the cleanest possible study of how the agent addresses them, as opposed to allowing them to occur naturally due to unintentional guide ambiguities.

Three of the four defects were resolved before the final stage of model development. Of these three, two were resolved without human intervention (Cases 1 and 4), and one was addressed only after a hint from the human-in-the-loop (Case 2). The fourth was never detected (Case 3). Detection tracked whether the guide contradicted itself, not the severity or even the visibility of the symptom. Case 1\textquotesingle s sign reversal made the boundary-value problem unsolvable, and the sweep failed outright. Case 2 made the Jacobian structurally singular. Case 4 produced transport coefficients thirty-fold higher than their correct values and the agent rejected them on magnitude at Stage 3. In principle, Case 2 should have been the most consequential (breaking a conservation law) and was also the most conspicuous, but a visible symptom alone was not sufficient to be detected by the agent. Such errors often required a second textual signal, for instance, the modified guide contradicting itself. Case 1\textquotesingle s reversed flux had the opposite sign to the Nernst--Planck boundary condition stated elsewhere in the same document, and Case 4\textquotesingle s added units statement was incompatible with the Fuller--Schettler--Giddings correlation\textquotesingle s own prefactor. Both errors were caught unaided. Case 2 left no such contradiction accessible to the agent and thus needed a hint from the operator.

Of the four planted errors, Case 3 (omission of porous media correction for diffusivities) is the most informative failure, and the one that establishes the limits of the agents' diagnostic capability. It is a plain constitutive error, no subtler in isolation than the others, but it was the only defect propagated consistently through every statement, example, and derived number in the guide. It was also numerically silent: the residuals converged at every stage, global charge conservation held to 2.4 × 10\textsuperscript{--11}, and the one diagnostic that did register it, a liquid-potential drop at the bottom edge of the expected band, was explained away using the model\textquotesingle s own inflated conductivity. Because the error could be compensated with the choice of kinetic parameters, the fitting step absorbed most of it, and the delivered model still reproduced the experimental polarization while carrying an exchange current density 41\% below every other build. This is the error class that staged construction does not close on its own and requires discerning human rigor. This class of error is also quite common and is not specific to this multiphysics problem. A wrong constitutive relation transcribed faithfully and absorbed by a fit is a hazard of any model-building. The remedy is external reference, such as an independent implementation, an analytical limit, or a measurement the fit was not trained on. In this case, having the COMSOL model as an independent, orthogonal reference allowed the team to resolve silent errors in the implementation guide that caused deviations between the guide and the COMSOL implementation, which was treated as a ``ground truth'' or golden model.

The study also shows that the build process is resilient to minor errors that would otherwise have major impacts. Beyond syntax errors, the agents caught common mistakes that hinder model development and are tedious to find manually, including sign, unit, and math errors. Human mistakes are always a possibility regardless of experience level, so agents that can dissect code and dozens of parameters and equations in a fraction of the time provide a significant safety net. All syntax errors and most convergence errors arising from code implementation were caught by autonomous agent review, without human input.

\subsection{Fully autonomous builds}\label{ms:fully-autonomous-builds}

\begin{figure}[b!]
\centering
\includegraphics[width=\textwidth]{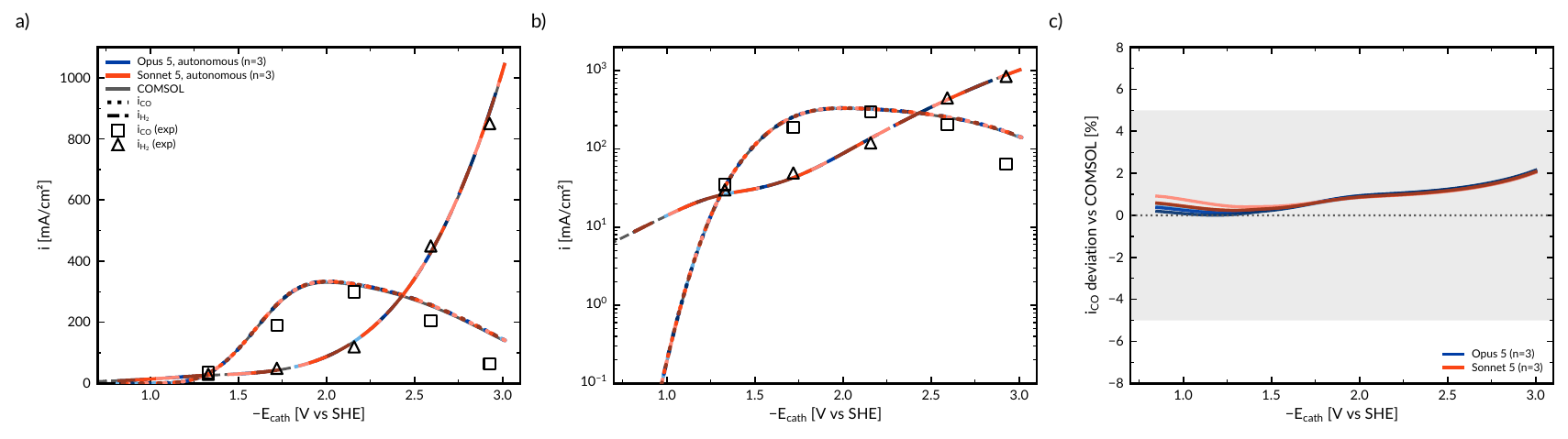}
\caption{\textbf{Autonomous agent builds without a human-in-the-loop.} Comparison of Claude Opus 5 and Sonnet 5 build replicate \textbf{(a)} linear-scale and \textbf{(b)} log-scale polarization curves to COMSOL and experimental data. \textbf{(c)} Deviation of each build\textquotesingle s \emph{i\textsubscript{CO}} from COMSOL at the COMSOL sweep potentials.}\label{fig:6}
\end{figure}

With human review eventually finding nothing else to catch during builds, the next step was pushing the bounds of agent autonomy. We gave the finalized implementation guide and the same prompt to three separate instances each of Claude Opus 5 and Sonnet 5 and instructed them to bypass the stop gates completely for a fully autonomous build with no human input. The agents were instructed to not refer to anything other than the guide, isolating the autonomous agents from previous builds and COMSOL data. The agent's project memory, which carried a few results of earlier builds, was active in these builds as in all others (Methods); a tenth autonomous build made without it fitted \emph{i\textsubscript{0}} = 106.59 A m\textsuperscript{−2} and λ = 1.6109 eV, inside the range of all nine builds. They checked their own plots and validated each stage\textquotesingle s output against the physical and numerical checks provided in the guide. All six builds produced virtually the same output (\textbf{\hyperref[fig:6]{Fig. 6} and Supplementary Note 8),} replicating the canonical model with CO\textsubscript{2}R fitted parameters of \emph{i\textsubscript{0}} = 106.56 ± 0.07 A m\textsuperscript{−2} and λ = 1.6113 ± 0.0002 eV for Opus builds and \emph{i\textsubscript{0}} = 106.51 ± 0.01 A m\textsuperscript{−2} and λ = 1.6108 ± 0.0003 eV for Sonnet builds (Supplementary Table 9).

These results show that LLM-assisted model building can culminate in fully autonomous builds once the implementation guide is fully developed. We emphasize that this was achievable only after substantial guide iteration. Nonetheless, the success of the Sonnet builds demonstrates a major payoff of the refinement, which is that less powerful models can do the same task at a cheaper rate. Usage costs can limit the ability of users to perform complex or large tasks within a certain time frame, but with well-defined tooling and harnesses (in our case, the specification), the scope of ambition can be made larger with cheaper agents. For instance, autonomy allows agents to run replicates or comparative studies in parallel without constant supervision, yet it can lead to fast usage depletion or high costs if used with stronger models like Opus and Fable. Cheaper agents, like Sonnet, require a fraction of the cost and can yield similar results depending on the task. While achieving autonomy for model building is a great milestone, it does not eliminate the risk that an autonomous build can silently inherit whatever is wrong with its specification guide, once again emphasizing the importance of the human in establishing a stable implementation guide to harness the multiphysics agent.

\section{Discussion}\label{ms:discussion}

This work extends LLM-assisted scientific computing beyond benchmark PDEs and weakly coupled or single-physics problems to a true multiphysics model of an experimentally relevant electrochemical device. Although the model is one-dimensional, it is not numerically simple: Nernst--Planck transport, carbonate buffer chemistry, electrochemical kinetics, Stefan--Maxwell gas transport, Darcy flow, and electrostatics are coupled through nonlinear source terms and transport coefficients in a stiff system containing 1,480 degrees of freedom. The resulting model reproduces expected physics and was calibrated to real experimental polarization and selectivity data and verified against an independently implemented COMSOL model. That provides a substantially stronger test of whether an agent can construct a model whose coupled physics actually reproduce the behavior of a real electrochemical system. More broadly, the results show that LLM-assisted multiphysics modeling can shift much of the technical burden from numerical implementation to physical specification and verification. The researcher still has to define the governing physics, choose the constitutive relationships and boundary conditions, recognize unphysical behavior, and validate the model. These decisions are where domain knowledge enters the model. For instance, we describe the porous electrode as a macrohomogeneous medium rather than a pore-resolved domain because that description resolves the polarization, selectivity, and concentration profiles studied here. The agent can implement either description, but the assumptions and simplifications appropriate to the question being asked remain the choice of the researcher. What the agent removes is the requirement that the researcher also translate that understanding into specialized numerical code.

The implementation guide therefore becomes the central scientific artifact in this workflow. \textbf{Supplementary Table 12} summarizes the components that most strongly determined whether independent agents built the intended model rather than a plausible approximation, including complete governing equations and boundary conditions, explicit unit, sign, and reference-frame conventions, numerical methods, staged verification criteria, and expected physical magnitudes. These requirements are not simply instructions for an AI agent. Together, they provide an explicit record of the assumptions, numerics, and physical reasoning underlying the model that can be inspected, revised, and shared independently of any particular implementation. The limitations are equally important: staged verification catches many implementation and convergence failures, but a physically incorrect assumption stated consistently throughout the specification can still be reproduced faithfully and even obscured by parameter fitting. Independent builds, external references, and human physical intuition therefore remain important safeguards.

The opportunity is therefore not to remove the researcher from continuum modeling, but to make the researcher's understanding of the physics the primary interface to it\textbf{.} The models produced here are open source and fully inspectable, and the agent that builds them can trace numerical behavior back through residuals, local concentrations, governing equations, and physical mechanisms in the same language the researcher uses to reason about the system. Researchers who understand an electrochemical device but are not specialists in numerical methods can therefore build, troubleshoot, extend, and share rigorous multiphysics models without surrendering control of either the physics or the implementation. By replacing a barrier of specialized code and proprietary software with one of stating and evaluating the physics precisely, agentic AI can make transparent multiphysics modeling accessible to a much broader scientific community, expanding the role of multiphysics models in the future of science and engineering research.

\section{Methods}\label{ms:methods}

\subsection{Model system and governing equations}\label{ms:model-system-and-governing-equations}

The model developed herein is based upon prior work on modeling electrochemical CO\textsubscript{2} reduction in 1D and, importantly, presents a stiff, highly nonlinear and coupled testbed with which to evaluate the efficacy and capabilities of our agentic multiphysics approach. Evaluating the effectiveness of building highly coupled and stiff numerical models with LLM agents requires an existing solution against which to verify the implementation. We therefore rebuilt a CO\textsubscript{2}R porous electrode model assembled from three previous COMSOL-based studies (\textbf{\hyperref[fig:1]{Fig. 1}b}). The domain structure and GDE formulation follow Weng et al.\textsuperscript{\msc{7}}; the Marcus--Hush--Chidsey (MHC) kinetics\textsuperscript{\msc{27}} follow Lees et al.\textsuperscript{\msc{36}}; and the Ag-catalyzed CO\textsubscript{2}R model of Muhieddine Orfali et al.\textsuperscript{\msc{37}} serves as the COMSOL reference that we rebuilt for benchmarking. We resolve the cathode catalyst layer (CL) and gas diffusion layer (GDL) only and do not model the anode or membrane. The cathode alone is a sufficiently demanding multiphysics problem for a first demonstration, and full-cell models and models beyond 1D will be subject of future work.

This system possesses each of the couplings described above. Two electrochemical reactions proceed at the catalyst surface,
\begin{equation}\text{CO\textsubscript{2} + H\textsubscript{2}O + 2e\textsuperscript{--} → CO + 2OH\textsuperscript{--}}\label{eq:ms1}\end{equation}
\begin{equation}\text{2H\textsubscript{2}O + 2e\textsuperscript{--} → H\textsubscript{2} + 2OH\textsuperscript{--}}\label{eq:ms2}\end{equation}
and both generate hydroxide, which is consumed by the carbonate buffer equilibria in the electrolyte,
\begin{equation}\text{CO\textsubscript{2} + OH\textsuperscript{--} ⇌ HCO\textsubscript{3}\textsuperscript{--}}\label{eq:ms3}\end{equation}
\begin{equation}\text{HCO\textsubscript{3}\textsuperscript{--} + OH\textsuperscript{--} ⇌ CO\textsubscript{3}\textsuperscript{2--} + H\textsubscript{2}O}\label{eq:ms4}\end{equation}
closing a loop in which current raises the local pH, the elevated pH consumes dissolved CO\textsubscript{2}, and the resulting CO\textsubscript{2} depletion suppresses the kinetic CO\textsubscript{2}R rate that produced the current as per mass transport limitations. Liquid-phase transport is described by Nernst--Planck dilute-solution theory with acid--base and electrochemical source terms, gas transport by Stefan--Maxwell concentrated-solution theory with Henry's law phase transfer, and gas-phase momentum by Darcy's law; CO\textsubscript{2}R is modeled with MHC kinetics, evaluated with the closed-form approximation\textsuperscript{\msc{38}}, and the competing hydrogen evolution reaction (HER) with Butler--Volmer kinetics. The model is one-dimensional, steady-state, and isothermal at 50 °C, with fixed porosity and liquid saturation. The CL is treated as three phases, solid catalyst, ionomer/liquid, and gas-filled pore, each with fixed volume fractions defined by experimentally measured porous electrode structure. The GDL is treated as dry, and crossover to the anolyte is not modeled. The complete governing equations and boundary conditions are given in Supplementary Notes 1 and 2, the parameter set in Supplementary Note 9 and Supplementary Table 11, and the COMSOL build used for benchmarking in Supplementary Note 3.

\subsection{The implementation guide}\label{ms:the-implementation-guide}

Critical to building well-developed, rigorous models are equally well-developed implementation guides for the agent. Implementation guides are machine-readable specifications (often in the form of markdown files) that fully describe and specify the problem to be modeled, including all relevant parameters, equations, boundary conditions, and include a scaffolded build plan that the coding agent must follow. Essentially, the implementation guide is a human-specified harness to constrain the multiphysics agent. These guides link the science from the human to the numerical implementation of the agent. Ensuring these guides are thorough and precise is necessary to prevent ambiguity for the agent that might lead to errors and inconsistencies in building the model. Hence, the human operator's understanding of the physics and chemistry of a given multiphysics problem remains crucial to developing the model. In this workflow, the agent's primary role is not inference but implementation, removing the numerical implementation barrier that otherwise hinders scientific modeling and has been largely inaccessible to chemists in the past.

The guide used in this case study includes 18 sections covering everything needed for the full implementation of the multiphysics model: all equations, parameters, and solver information. Key structural elements include the complete physics of the system, full parameter tables, the numerical methods, and anticipated failure modes with solutions (Supplementary Note 10). The complete guide is available in the repository given under Code availability and is also provided as a supplementary file.

In all, the guide\textquotesingle s components are made to be as explicit and precise as possible to mitigate any inference or hallucination on the agent\textquotesingle s part. This mirrors the prompting literature, where explicit, structured, example-bearing instructions outperform terse ones\textsuperscript{\msc{39}}. Importantly, a proper specification can also be read and audited by a human such that the human's physical intuition properly guides the development of the guide and constrains the multiphysics problem. The physics lay out the governing equations explicitly, not citing by name such as ``standard Nernst--Planck''. The parameter tables list every constant, including its name, symbol, value, units, and source. The numerical methods in this study include a finite-volume discretization, using a Newton solver with damping (Supplementary Table 5). Bootstrap protocols for preconditioning and methods for initializing solvers are included to improve convergence for stiff problems such as the CO\textsubscript{2}R problem employed herein.

The methods were specified from physical insight and user experience building models in COMSOL, described in sufficient detail to prevent the agent from making any critical algorithmic decisions. We note that the significant degree of scaffolding used in this workflow is required for the current state of frontier models due to their tendency to invoke unphysical shortcuts or to hallucinate parameters\textsuperscript{\msc{40}}. However, purpose-built agents\textsuperscript{\msc{41},\msc{42}} or future frontier models may be able to handle greater flexibility in specification. Notably, this methodology is also transferable to other systems: a fuel cell or a Li-ion pseudo-two-dimensional model\textsuperscript{\msc{43},\msc{44}} guide would follow the same structure, with only the equations and parameters needing changes. Regardless, no specification is immune to ambiguity, which is why staged verification and guide refinement are essential. Having a human-in-the-loop for the development of the guide, at this stage, appears essential.

Because the guide is a plain-text markdown file, it can be version-controlled and developed collaboratively. In this work, the implementation guide was maintained in a shared GitHub repository, where co-authors iteratively refined its physics, parameters, and numerical methods and shared the next refinement targets. The same versioned guide could be given to multiple researchers who independently instructed agents to build the model, where each independent model build was treated as a replicate. Divergent outputs exposed ambiguities or errors in the guide (Supplementary Tables 8 and 10), while agreement across independent builds gave confidence in both the model and the specification. This ability to collaborate on a single human- and machine-readable artifact, rather than on an opaque, version-locked solver project, is itself a central advantage of the workflow.

\subsection{Staged construction and stop gates}\label{ms:staged-construction-and-stop-gates}

Building the model in stages of sequentially increasing complexity is essential to enable human review of output, identify and resolve issues before they accumulate and compound, and subsequently iterate upon the guide (\hyperref[fig:1]{\textbf{Fig. 1}}). The guide decomposes the model into five stages of increasing physical complexity. In Stage 1, liquid Nernst--Planck, buffer reactions, and electrochemical kinetics are included with Dirichlet boundary conditions (BCs), initializing the problem with 400 degrees of freedom (DOFs) at the production mesh of 80 CL cells and 120 GDL cells. In Stage 2, Sherwood--Reynolds mass-transfer (Robin) BCs replace the Dirichlet to capture finite mass transfer at the electrolyte boundary. In Stage 3, Stefan--Maxwell gas transport and Henry\textquotesingle s law phase transfer are added, expanding the system to 1,200 DOFs. In Stage 4, the Darcy pressure equation is finalized, completing the gas-phase momentum balance. A fitting step then estimates the kinetic parameters for CO\textsubscript{2}R and HER against the experimental data of Romiluyi et al.\textsuperscript{\msc{25}}, exploiting the well-conditioned Stage 4 residual before the coupling of the charged species to electrostatic potential causes stiffness to grow further. Stage 5 then turns on the full electrochemistry for the migration term in the Nernst--Planck flux, solving for liquid potential φ\textsubscript{L} monolithically from the per-cell ionic charge balance in the CL and solid potential φ\textsubscript{s} in the GDL and CL via Ohm\textquotesingle s law. This brings the full system to 1,480 DOFs at the 80-CL/120-GDL cell production mesh (Supplementary Table 2). The kinetic parameters fitted at Stage 4 are held fixed in Stage 5, where migration raises the simulated CO partial current (\hyperref[fig:2]{\textbf{Fig. 2}}a).

Crucially, the stages are designed to include stop gates where the human-in-the-loop can review quantitative physicality checks (\emph{e.g.,} checking electroneutrality to machine precision, mass conservation, current continuity, and Newton residual convergence) before authorizing the next stage. These checks are prescribed by the guide itself rather than left to the discretion of the agent. For each stage, the guide specifies the quantitative pass/fail criteria as well as diagnostic plots the agent must produce (Supplementary Note 5), to enable criteria-driven and reproducible human review across builders. These output plots force data visualization, allowing the user to directly evaluate the model and accelerate diagnostics. In our case, a physical understanding of the expected general trends, including those in polarization behavior, faradaic efficiencies (FEs), and concentration profiles, makes these plots more interpretable than a list of numbers. A visual inspection to verify that trends are accurate, such as pH increasing with cathodic current and CO\textsubscript{2} concentration dropping with increasing overpotential, provides evidence that the model is building properly.

This framework is essential for revealing failure modes in isolation and facilitating independent human verification of all model outputs, rather than relying solely on the agent's self-assessment. For instance, an error in the buffer equilibria implementation affects local pH, and thus CO\textsubscript{2} concentration, reaction rates, liquid phase potential, etc., contaminating practically every part of the highly coupled problem. Locating errors in a liquid-only 400-DOF sub-problem is much simpler than allowing those same errors to get buried inside the full-physics 1,480-DOF system. Establishing stop gates addresses the issue of compounding complexity and allows the user to more effectively troubleshoot. Errors during model building could be due to mistakes made by the agent, but these stop gates also reveal inconsistencies or ambiguities in the guide, allowing the user to refine and iterate to make it more comprehensive and reproducible across instances of the agent. We note that frontier AI models are improving rapidly, even within the timespan of this study, and could eventually check their own work or work in teams of agents\textsuperscript{\msc{18},\msc{19}} to develop these rigorous models. At this stage, however, we find that building a model specification of this scale from scratch in one prompt is not practical, and scaffolding makes the workflow reliable and generalizable.

\subsection{Agents and models}\label{ms:agents-and-models}

Throughout guide development, we used the most capable Claude model available at the time because of the task\textquotesingle s complexity. In earlier builds from less mature guides, weaker Sonnet-tier models introduced more errors than Opus, and the additional time spent correcting them, at comparable token costs, outweighed any benefit. Opus 4.6 was used to initialize guide development and perform much of the front-end refinement. Opus 4.7 was then primarily used for finalizing the model, and Opus 5 was used to make the final builds for comparison to COMSOL. Opus 4.8 was used for the planted-error study, and Opus 5 and Sonnet 5 for the fully autonomous builds. The choice to use a frontier model (Claude) rather than training our own purpose-built agent was made to first demonstrate the feasibility of the methodology with a well-established and powerful agent that, in theory, possesses intrinsic knowledge of solving similar nonlinear PDE problems.

These agents were used by multiple users across laboratories, with varying experience levels, to develop the same implementation guide. When building models, researchers tracked their build processes in build dialogue files (Supplementary Note 7 and Supplementary Table 8) and instructed their agents to refer only to the implementation guide (apart from the agent's project memory, described below), so that its quality could be evaluated on its own and without bias from other builds or documents in the repository. In practice, more resources and context are beneficial for agents, and comparing replicates built from the same guide proved helpful in compensating for the intrinsic probabilistic nature of LLMs.

The coding agent's persistent project memory was active in all nine builds reported here. Its index, loaded at the start of each session, quoted results of builds made from an earlier version of the guide, including \emph{i\textsubscript{0}} ≈ 106.5 A m\textsuperscript{−2}, a Stage-5 peak \emph{i\textsubscript{CO}} ≈ 334 mA cm\textsuperscript{−2} and a note that a COMSOL run at that \emph{i\textsubscript{0}} agreed to about 1\%. It held no value of λ or the HER parameters, no polarization curve and no COMSOL output, and no build opened another build's files or COMSOL output before it was complete. The guide itself states expected values at its stop gates, about twenty of them measured on the first of the nine builds. The agreement between builds does not rest on these values: λ, the HER parameters and the full 89-point polarization curves, which had no stated target, agree across builds as closely as \emph{i\textsubscript{0}} does, and each fit starts from the guide's prescribed initial point and is deterministic. A tenth build made in a directory containing only the guide, with no project memory, fitted \emph{i\textsubscript{0}} = 106.59 A m\textsuperscript{−2} and λ = 1.6109 eV (Supplementary Note 8).

\subsection{Planted-error study}\label{ms:planted-error-study}

To perform the planted error study, specific defects were planted in the specification guide, not in the code, reproducing the failure mode of practical interest. In this failure mode, the statement of the physics handed to the implementer is itself wrong and is transcribed faithfully, propagating the error from the implementation guide to the codebase. Detection therefore requires the agent to recognize a conflict between what it has been told to implement and either the physics or the remainder of the specification and to raise that conflict rather than deviate silently. Each build received the same instruction: implement the model exactly as the guide specifies, consult no other file, halt at each stage gate for human review, and maintain a running build log recording every error encountered, every error corrected without human input, and every discrepancy noticed in the guide. The agent was not told that a defect existed or where. At each stage gate, the human reviewer inspected the plotted stage output and approved or rejected progression without auditing the source code. A defect was scored as detected only if the build log identified it as a defect, named the affected relation, and proposed a correction. Where a modified guide contained explanatory or diagnostic text naming the affected quantity, that text was edited along with the defect, so detection could not follow from an isolated self-contradiction. Per-case results are given in Supplementary Note 6 and Supplementary Table 7.

\subsection{Solve-time benchmarking and mesh convergence}\label{ms:solve-time-benchmarking-and-mesh-convergence}

Both implementations were timed on the same workstation (Intel Xeon w7-3565X, 32 cores/64 threads, 191 GB RAM) at matched core budget, with COMSOL on 32 cores (20,033 DOFs; Supplementary Table 1) and Julia on 32 threads (20,038 DOFs; Supplementary Note 4 and Supplementary Table 6). COMSOL swept 61 potentials at 50 mV spacing and Julia 89 at 25 mV, and both converged at every potential (\textbf{\hyperref[fig:4]{Fig. 4}a--b}). Reported Julia times are the fastest of three repetitions, with run-to-run variability reported in Supplementary Table 4, and the COMSOL time was the same in both of its recorded runs; process start-up and just-in-time compilation, a further 15--20 s per invocation, are excluded from both. The Julia data were obtained under the Windows Subsystem for Linux rather than natively, because the MUMPS binary distribution, used for the solver ablation, has no working Windows-native build. Mesh convergence of the Stage 5 solution is documented in Supplementary Note 4 and Supplementary Fig. 2.

\subsection{Reporting summary}\label{ms:reporting-summary}

Further information on research design is available in the Nature Portfolio Reporting Summary linked to this article.

\section{Data availability}\label{ms:data-availability}

The experimental partial current densities used for kinetic parameter estimation are taken from Romiluyi et al. and are reproduced in Supplementary Table 3. The simulated data underlying Figs. 2, 3 and 6, and the COMSOL results they are compared with, are included in the code repository (see Code availability); the timings in Fig. 4 are listed in Supplementary Tables 4 and 6.

\section{Code availability}\label{ms:code-availability}

The implementation guide, the Julia model source for all five build stages, and the build dialogue logs are available at \url{https://github.com/Sebobbit/CO2R-Bulk-Scale-Model}, with the Julia source released under the MIT license. The four modified guides used in the planted-error study, together with the unmodified control guide, are provided as separate files in the same repository. The COMSOL reference model is included; opening and running it requires a COMSOL Multiphysics license (v6.4) with the Electrochemistry and CFD modules, and its exported results are provided as plain-text files so that the comparison in Fig. 3 can be reproduced without a license.

\section{Acknowledgements}\label{ms:acknowledgements}

The authors acknowledge professors Miguel A. Modestino and Rohini B. Chandran for helpful discussions.

\section{Funding}\label{ms:funding}

S.C. discloses support for the research of this work from the National Science Foundation Graduate Research Fellowship Program {[}DGE-2234660{]}. J.C.B. discloses support for the research of this work from Schmidt Science Fellows, in partnership with the Rhodes Trust. E.W.L. acknowledges funding support from the Peter Wall Legacy Award and the Natural Sciences and Engineering Research Council of Canada Discovery Grants program (RGPIN-2025-04804). Any opinions, findings, and conclusions or recommendations expressed in this material are those of the authors and do not necessarily reflect the views of the National Science Foundation.

\section{Author contributions}\label{ms:author-contributions}

S.C.: Investigation, Visualization, Writing -- original draft, Writing -- review \& editing. M.F.S.: Investigation, Software. S.M.: Investigation. E.W.L.: Investigation, Supervision, Writing -- review \& editing. J.C.B.: Investigation, Supervision, Writing -- review \& editing.

\section{Competing interests}\label{ms:competing-interests}

J.C.B., S.C., M.F.S. and E.W.L. are inventors on a provisional patent application filed by New York University and the University of British Columbia covering the specification-driven modeling workflow described here (US Provisional Application no. 64/168,700). The authors declare they have no other conflicts of interest.

\clearpage
\begin{msbibliography}
\bibitem{ms1} Chu, S. \& Majumdar, A. Opportunities and challenges for a sustainable energy future. \emph{Nature} \textbf{488}, 294--303 (2012).

\bibitem{ms2} De Luna, P. \emph{et al.} What would it take for renewably powered electrosynthesis to displace petrochemical processes? \emph{Science} \textbf{364}, eaav3506 (2019).

\bibitem{ms3} Burdyny, T. \& Smith, W. A. CO2 reduction on gas-diffusion electrodes and why catalytic performance must be assessed at commercially-relevant conditions. \emph{Energy Environ. Sci.} \textbf{12}, 1442--1453 (2019).

\bibitem{ms4} Welch, A. J. \emph{et al.} Operando Local pH Measurement within Gas Diffusion Electrodes Performing Electrochemical Carbon Dioxide Reduction. \emph{J. Phys. Chem. C} \textbf{125}, 20896--20904 (2021).

\bibitem{ms5} Bui, J. C. \emph{et al.} Continuum Modeling of Porous Electrodes for Electrochemical Synthesis. \emph{Chem. Rev.} \textbf{122}, 11022--11084 (2022).

\bibitem{ms6} Newman, J. \& Balsara, N. P. \emph{Electrochemical Systems.} (John Wiley \& Sons, Incorporated, Newark, 2021).

\bibitem{ms7} Weng, L.-C., Bell, A. T. \& Weber, A. Z. Modeling gas-diffusion electrodes for CO2 reduction. \emph{Phys. Chem. Chem. Phys.} \textbf{20}, 16973--16984 (2018).

\bibitem{ms8} Weng, L.-C., Bell, A. T. \& Weber, A. Z. Towards membrane-electrode assembly systems for CO2 reduction: a modeling study. \emph{Energy Environ. Sci.} \textbf{12}, 1950--1968 (2019).

\bibitem{ms9} Ringe, S. \emph{et al.} Understanding cation effects in electrochemical CO2 reduction. \emph{Energy Environ. Sci.} \textbf{12}, 3001--3014 (2019).

\bibitem{ms10} Newman, J. Numerical Solution of Coupled, Ordinary Differential Equations. \emph{Ind. Eng. Chem. Fundam.} \textbf{7}, 514--517 (1968).

\bibitem{ms11} Newman, J. S. \& Tobias, C. W. Theoretical Analysis of Current Distribution in Porous Electrodes. \emph{J. Electrochem. Soc.} \textbf{109}, 1183 (1962).

\bibitem{ms12} Dickinson, E. J. F., Ekström, H. \& Fontes, E. COMSOL Multiphysics®: Finite element software for electrochemical analysis. A mini-review. \emph{Electrochem. Commun.} \textbf{40}, 71--74 (2014).

\bibitem{ms13} Sulzer, V., Marquis, S. G., Timms, R., Robinson, M. \& Chapman, S. J. Python Battery Mathematical Modelling (PyBaMM). \emph{J. Open Res. Softw.} \textbf{9}, (2021).

\bibitem{ms14} Permann, C. J. \emph{et al.} MOOSE: Enabling massively parallel multiphysics simulation. \emph{SoftwareX} \textbf{11}, 100430 (2020).

\bibitem{ms15} Peng, R. D. Reproducible Research in Computational Science. \emph{Science} \textbf{334}, 1226--1227 (2011).

\bibitem{ms16} M. Bran, A. \emph{et al.} Augmenting large language models with chemistry tools. \emph{Nat. Mach. Intell.} \textbf{6}, 525--535 (2024).

\bibitem{ms17} Boiko, D. A., MacKnight, R., Kline, B. \& Gomes, G. Autonomous chemical research with large language models. \emph{Nature} \textbf{624}, 570--578 (2023).

\bibitem{ms18} Ni, B. \& Buehler, M. J. MechAgents: Large language model multi-agent collaborations can solve mechanics problems, generate new data, and integrate knowledge. \emph{Extreme Mech. Lett.} \textbf{67}, 102131 (2024).

\bibitem{ms19} Park, D., Moon, H. \& Ryu, S. A self-correcting multi-agent LLM framework for language-based physics simulation and explanation. \emph{Npj Artif. Intell.} \textbf{2}, 10 (2026).

\bibitem{ms20} Lupoiu, R. \emph{et al.} A multi-agentic framework for real-time, autonomous freeform metasurface design. \emph{Sci. Adv.} \textbf{11}, eadx8006 (2025).

\bibitem{ms21} Li, S. \emph{et al.} CodePDE: An Inference Framework for LLM-driven PDE Solver Generation. Preprint at \url{https://doi.org/10.48550/arXiv.2505.08783} (2026).

\bibitem{ms22} Adhikari, S., Noorsumar, G. \& Jensen, Ø. PDE-Agents: An LLM-Orchestrated Multi-Agent Framework for Automated Finite Element Simulations with Knowledge Graph-Augmented Reasoning. Preprint at \url{https://doi.org/10.48550/arXiv.2606.07850} (2026).

\bibitem{ms23} Bezanson, J., Edelman, A., Karpinski, S. \& Shah, V. B. Julia: A Fresh Approach to Numerical Computing. \emph{SIAM Rev.} \textbf{59}, 65--98 (2017).

\bibitem{ms24} Oberkampf, W. L. \& Trucano, T. G. Verification and validation in computational fluid dynamics. \emph{Prog. Aerosp. Sci.} \textbf{38}, 209--272 (2002).

\bibitem{ms25} Romiluyi, O., Danilovic, N., Bell, A. T. \& Weber, A. Z. Membrane-electrode assembly design parameters for optimal CO2 reduction. \emph{Electrochem. Sci. Adv.} \textbf{3}, e2100186 (2023).

\bibitem{ms26} Corpus, K. R. M. \emph{et al.} Coupling covariance matrix adaptation with continuum modeling for determination of kinetic parameters associated with electrochemical CO2 reduction. \emph{Joule} \textbf{7}, 1289--1307 (2023).

\bibitem{ms27} Chidsey, C. E. D. Free Energy and Temperature Dependence of Electron Transfer at the Metal-Electrolyte Interface. \emph{Science} \textbf{251}, 919--922 (1991).

\bibitem{ms28} Nelder, J. A. \& Mead, R. A Simplex Method for Function Minimization. \emph{Comput. J.} \textbf{7}, 308--313 (1965).

\bibitem{ms29} Mogensen, P. K. \& Riseth, A. N. Optim: A mathematical optimization package for Julia. \emph{J. Open Source Softw.} \textbf{3}, 615 (2018).

\bibitem{ms30} Schenk, O. \& Gärtner, K. Solving unsymmetric sparse systems of linear equations with PARDISO. \emph{Future Gener. Comput. Syst.} \textbf{20}, 475--487 (2004).

\bibitem{ms31} Davis, T. A. Algorithm 832: UMFPACK V4.3-\/-\/-an unsymmetric-pattern multifrontal method. \emph{ACM Trans. Math. Softw. TOMS} \textbf{30}, 196--199 (2004).

\bibitem{ms32} Amestoy, P. R., Duff, I. S., L'Excellent, J.-Y. \& Koster, J. A Fully Asynchronous Multifrontal Solver Using Distributed Dynamic Scheduling. \emph{SIAM J. Matrix Anal. Appl.} \textbf{23}, 15--41 (2001).

\bibitem{ms33} Revels, J., Lubin, M. \& Papamarkou, T. Forward-Mode Automatic Differentiation in Julia. Preprint at \url{https://doi.org/10.48550/arXiv.1607.07892} (2016).

\bibitem{ms34} Brady, N. W., Mees, M., Vereecken, P. M. \& Safari, M. Implementation of Dual Number Automatic Differentiation with John Newman's BAND Algorithm. \emph{J. Electrochem. Soc.} \textbf{168}, 113501 (2021).

\bibitem{ms35} Ouyang, S., Zhang, J. M., Harman, M. \& Wang, M. An Empirical Study of the Non-Determinism of ChatGPT in Code Generation. \emph{ACM Trans. Softw. Eng. Methodol.} \textbf{34}, 42:1-42:28 (2025).

\bibitem{ms36} Lees, E. W., Bui, J. C., Romiluyi, O., Bell, A. T. \& Weber, A. Z. Exploring CO2 reduction and crossover in membrane electrode assemblies. \emph{Nat. Chem. Eng.} \textbf{1}, 340--353 (2024).

\bibitem{ms37} Muhieddine Orfali, D. \emph{et al.} Pulsed Electrolysis Promotes Catalyst Activity in Dilute CO2 Streams. \emph{ACS Energy Lett.} \textbf{11}, 4988--4997 (2026).

\bibitem{ms38} Zeng, Y., Smith, R. B., Bai, P. \& Bazant, M. Z. Simple formula for Marcus--Hush--Chidsey kinetics. \emph{J. Electroanal. Chem.} \textbf{735}, 77--83 (2014).

\bibitem{ms39} Schulhoff, S. \emph{et al.} The Prompt Report: A Systematic Survey of Prompt Engineering Techniques. Preprint at \url{https://doi.org/10.48550/arXiv.2406.06608} (2025).

\bibitem{ms40} Huang, L. \emph{et al.} A Survey on Hallucination in Large Language Models: Principles, Taxonomy, Challenges, and Open Questions. \emph{ACM Trans. Inf. Syst.} \textbf{43}, 42:1-42:55 (2025).

\bibitem{ms41} Zou, Y. \emph{et al.} El Agente: An autonomous agent for quantum chemistry. \emph{Matter} \textbf{8}, 102263 (2025).

\bibitem{ms42} Zhang, Z. \emph{et al.} El Agente Forjador: Task-Driven Agent Generation for Quantum Simulation. Preprint at \url{https://doi.org/10.48550/arXiv.2604.14609} (2026).

\bibitem{ms43} Weber, A. Z. \& Newman, J. Modeling Transport in Polymer-Electrolyte Fuel Cells. \emph{Chem. Rev.} \textbf{104}, 4679--4726 (2004).

\bibitem{ms44} Doyle, M., Fuller, T. F. \& Newman, J. Modeling of Galvanostatic Charge and Discharge of the Lithium/Polymer/Insertion Cell. \emph{J. Electrochem. Soc.} \textbf{140}, 1526 (1993).
\end{msbibliography}